\documentclass[fleqn,usenatbib]{mnras}

\usepackage{newtxtext,newtxmath}
\usepackage{longtable}
\usepackage[T1]{fontenc}

\DeclareRobustCommand{\VAN}[3]{#2}
\let\VANthebibliography\thebibliography
\def\thebibliography{\DeclareRobustCommand{\VAN}[3]{##3}\VANthebibliography}

\usepackage{graphicx}	% Including figure files
\usepackage{amsmath}	% Advanced maths commands
\usepackage{pdflscape}
\usepackage{xcolor}
\newcommand{\apspr}{APSPR}  % needed by bibtex
\newcommand{\ergss}{\mathrm{erg} \ \mathrm{s}^{-1}}
\def\redbar#1{{\color{red}\rule{#1px}{8pt}}}
\def\bluebar#1{{\color{blue}\rule{#1px}{8pt}}}
\title[Long-Term X-Ray/UV Variability in ULXs]{Long-Term X-Ray/UV Variability in ULXs}

\author[N. Khan et al.]{
Norman Khan,$^{1}$\thanks{E-mail: nk7g14@soton.ac.uk}
Matthew. J. Middleton,$^{1}$
\\
$^{1}$School of Physics \& Astronomy, University of Southampton, Southampton, Southampton SO17 1BJ, UK\\
}

\date{Accepted XXX. Received YYY; in original form ZZZ}

\pubyear{2023}

\begin{document}
\label{firstpage}
\pagerange{\pageref{firstpage}--\pageref{lastpage}}
\maketitle

\begin{abstract}
The focus of NASA's \textit{Swift} telescope has been transients and
target-of-opportunity observing, resulting in many observations of
ultraluminous X-ray sources (ULXs) over the last $\sim 20$ years.  For the vast
majority of these observations, simultaneous data has been obtained using both
the X-ray telescope (XRT) and the ultraviolet and optical telescope (UVOT),
providing a unique opportunity to study coupled variability between these
bands. Using a sample of $\sim$40 ULXs with numerous repeat observations, we
extract stacked images to characterise the spatial extent of the UV-Optical
emission and extract long-term light curves to search for first-order linear
correlations between the UV and X-ray emission. We find that a small subset may
show weakly correlated joint variability, while other sources appear to display
non-linear relationships between the bands. We discuss these observations in
the context of several theoretical models: precession, irradiation of the outer
accretion disc and irradiation of the companion star.
We conclude that more complicated analysis or higher
quality data may be required to accurately constrain the nature of the joint
X-ray and UV/optical emission in these sources.
\end{abstract}

\begin{keywords}
stars: black holes -- stars: neutron
\end{keywords}

%%%%%%%%%%%%%%%%%%%%%%%%%%%%%%%%%%%%%%%%%%%%%%%%%%

%%%%%%%%%%%%%%%%% BODY OF PAPER %%%%%%%%%%%%%%%%%%

\section{Introduction}

Ultraluminous X-ray sources (ULXs) are defined as off-nuclear, point-like
sources with inferred isotropic luminosities exceeding $L > 10^{39} \ergss$
(see the reviews of \citealt{2007_Roberts_Ap&SS.311..203R,
2017_Kaaret_ARA&A..55..303K, 2023NewAR..9601672K}). Pulsations unambiguously
indicating the presence of neutron star (NS) accretors have been located in
several ULXs (\citealt{2014_Bachetti_Natur.514..202B,
2016_Furst_ApJ...831L..14F, 2017_Israel_Sci...355..817I,
2017_Tsygankov_A&A...605A..39T, 2018_Doroshenko_A&A...613A..19D,
2018_Carpano_MNRAS.476L..45C, 2019_Sathyaprakash_MNRAS.488L..35S,
2020_Rodriguez_ApJ...895...60R}) whilst the presence of a cyclotron resonant
scattering feature in M51 ULX-8 \citep{2018_Brightman_NatAs...2..312B} and
possible detection in NGC300 ULX1 \citep{2018_Walton_ApJ...857L...3W} also
suggest the presence of neutron stars. Given the luminosities of these sources
and the known accretor mass, it is inescapable that the accretion flow in such
systems is super-critical, either in the flow or onto the star itself.  

Should the Eddington limit be reached in the disc, then standard theory (i.e. \citealt{1973_Shakura_A&A....24..337S, 2007_Poutanen_MNRAS.377.1187P}) would
predict powerful, mass-loaded winds which have been located in several ULXs
to-date, both as imprints in the CCD spectra
(\citealt{2014_Middleton_MNRAS.438L..51M, 2015_Middleton_MNRAS.454.3134M,
2016_Walton_ApJ...826L..26W}) and using higher energy-resolution detectors
(\citealt{2016_Pinto_Natur.533...64P, 2017_Pinto_MNRAS.468.2865P,
2018_Kosec_MNRAS.473.5680K, 2021_Kosec_MNRAS.508.3569K}. A corollary of such
accretion flows is that, providing the wind is optically thick, there should be
some degree of collimation and the assumption of isotropy breaks down
\citep{2001_King_ApJ...552L.109K}. The resulting spectrum (and timing
properties) of a given ULX then depends on both the accretion rate and
inclination of the source \citep{2007_Poutanen_MNRAS.377.1187P,
2015_Middleton_MNRAS.447.3243M}.

Should the accretion rate be high, we would naturally expect high inclination
ULXs to be dim in the X-rays and instead peak at lower frequencies
\citep{2007_Poutanen_MNRAS.377.1187P}. With emission from the wind photosphere
being $\sim$Eddington, these may be prime candidates for detection by
next-generation deep surveys (e.g. LSST: Middleton et al. in prep). A prime
example of such an edge-on ULX is the Galactic source SS433 ($i=78^{\circ}$),
which, despite having an X-ray luminosity of only $\sim 10^{36} \ \ergss$
appears to share many of the same characteristics of ULXs and is inferred to
have considerably brighter, face-on X-ray luminosities
\citep{2002_Cherepashchuk_SSRv..102...23C, 2004_Fabrika_ASPRv..12....1F,
2016_Khabibullin_MNRAS.457.3963K, 2015_Liu_Natur.528..108L,
2021_Middleton_MNRAS.506.1045M} and emit at $\sim 10^{40} \ \ergss$ in the UV
\citep{1997_Dolan_A&A...327..648D}.

A change in the intrinsic X-ray/UV emission from a given ULX may be driven by
either a change in mass accretion rate at larger radius or a change in our line-of-sight
inclination to the inner-regions. The former may be the result of mass loss at
large radii \citep{2022_Middleton_MNRAS.509.1119M}, whilst the latter may be a
result of disc warping due to irradiation
(\citealt{1996_Pringle_MNRAS.281..357P}; \citealt{2013_Pasham_ApJ...774L..16P})
or precession of the super-critical disc and wind
(\citealt{2013_Pasham_ApJ...764...93P,2018_Middleton_MNRAS.475..154M,
2019_Middleton_MNRAS.489..282M}).

Whilst the intrinsic emission from high inclination ULXs may peak at low
frequencies, in the optical/UV band there will also be emission from the
secondary star which can be amplified if effectively irradiated, as well as
emission from the outer disc (again, if effectively irradiated). Whilst there
has been a great deal of study of ULXs in the optical -- specifically to
elucidate the nature of the companion star \citep{2014_Heida_MNRAS.442.1054H}
-- there has been limited exploration in the UV bands. However, it has been
observed that in one ULX, the UV emission appears extremely bright ($> 10^{39}$
erg/s: \citealt{2017_Kaaret_ARA&A..55..303K}) and in the case of the PULX, NGC
7793 P13, an optical/UV super-orbital period is seen, out of phase with the
X-ray super-orbital period \citep{2016_Furst_ApJ...831L..14F}. Although not
well explored, the correlation between X-ray and low frequency emission should
provide invaluable insight into the geometry and nature of the accretion flow.
In this paper, we explore the general shapes of the X-ray/UV correlations we
might expect for various scenarios and search for these within the Swift
lightcurves of several prominent ULXs.  

\section{Predictions}
\begin{figure}
    \centering
    \includegraphics[width=\columnwidth]{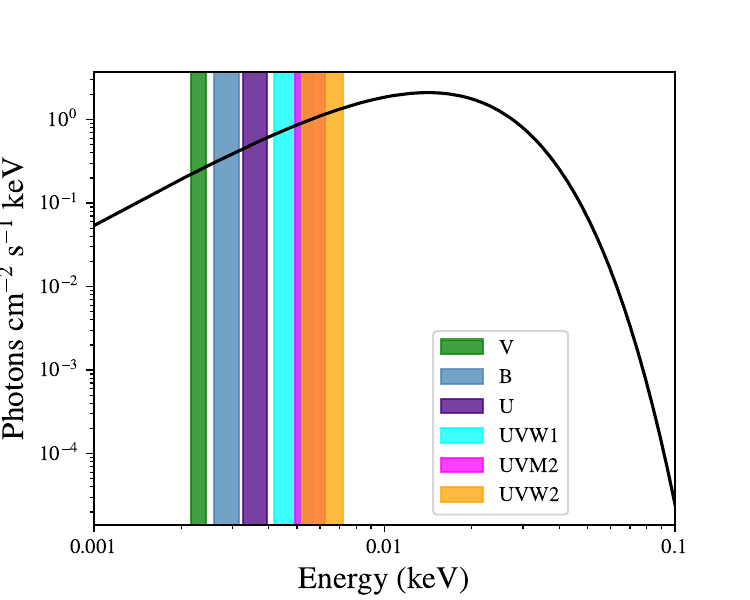}
    \caption[Blackbody spectrum with UVOT bands]{Black body spectrum
    (\texttt{bbody} in \texttt{xspec}) set at $kT = 0.005 \ \mathrm{keV}$.
    Coloured are the effective energy ranges of the UVOT bands
    used for the flux calculation
    shown in figure \ref{fig:mdot_vs_flux}.}
    \label{fig:blackbody_uvot}
\end{figure}

In this section, we consider the theoretical impact on the coupled X-ray/UV
properties of ULXs when ascribing the low frequency emission to various
locations in the system. To simplify this picture we make the explicit
assumption that ULXs which contain neutron stars all have dipole fields weak enough (or accretion rates
high enough) that the classical super-critical picture of disc accretion
\citep{2007_Poutanen_MNRAS.377.1187P} holds (i.e. that the magnetospheric
radius, $R_{\mathrm{M}}$, is much smaller than the spherisation radius, $R_{\mathrm{sph}}$).
Currently, measurements only exist for a handful of pulsating ULXs (see \citealt{2018_Walton_ApJ...857L...3W, 2018_Brightman_NatAs...2..312B, 2019_Middleton_MNRAS.486....2M, 2023_Furst_A&A...672A.140F}) but suggest field strengths of order $\sim 10^{11} - 10^{13} \ \mathrm{G}$
which according to the canonical picture of PULXs should tend to lead to the condition
that $R_{\mathrm{M}} < R_{\mathrm{sph}}$ \citep{2020_King_MNRAS.494.3611K}.
\subsection{Emission from the outer wind photosphere}
Following the super-critical model of \citealt{2007_Poutanen_MNRAS.377.1187P},
we assume that the wind photosphere extends out to some radius $r_{\rm out}$
and reprocesses the flux from the disc below the wind where the inflow starts
to become locally super-critical, and some fraction of the radiation produced
interior to $r_{\rm out}$ (the latter with a luminosity greater than or equal
to the Eddington luminosity). A full understanding of the shape and intensity
of the emergent SED requires full radiative magnetohydrodynamic (RMHD) simulations and extensive
post-processing, which has not yet been performed in the case of ULXs (although
see the work by \citealt{2017_Narayan_MNRAS.469.2997N} and
\citealt{2018_Lixin_ApJ...859L..20D}).  In the absence of numerical studies,
simple qualitative predictions are still possible. For a fixed inclination,
increasing the mass accretion rate pushes $r_{\rm out}$ to a larger radius due
to increased mass loading of the wind and the larger radial location of
$r_{\mathrm{sph}}$ \citep{2007_Poutanen_MNRAS.377.1187P}. Should the opening
angle of the wind be connected to the accretion rate at large radius (as it
would seem to be cf \citealt{2014_Jiang_ApJ...796..106J,
2019_Jiang_ApJ...880...67J}), then an increase in mass accretion rate will
increasingly collimate the X-ray emission from within.  What follows depends on
the orientation of the observer.  Should the observer be able to see into the
wind cone, then the X-ray luminosity at all energies will increase, and the
characteristic temperature associated with the spherization radius,
$T_{\mathrm{sph}}$ will decrease. The expansion of the wind photosphere to
larger radius will lead to a reduction in its temperature according to the
formula of \citealt{2007_Poutanen_MNRAS.377.1187P} (eq \ref{eq:T_sph}): 

\begin{equation}
	T_{\mathrm{sph}} \approx 0.8 f_{\mathrm{col}} \left(\frac{\zeta\beta}{\epsilon_{\mathrm{w}}}\right)^{1/2} m^{-1/4} \dot{m_{0}}^{-3/4} \ \mathrm{keV}
	\label{eq:T_sph}
\end{equation}

\begin{figure}
    \centering
    \includegraphics[width=\columnwidth]{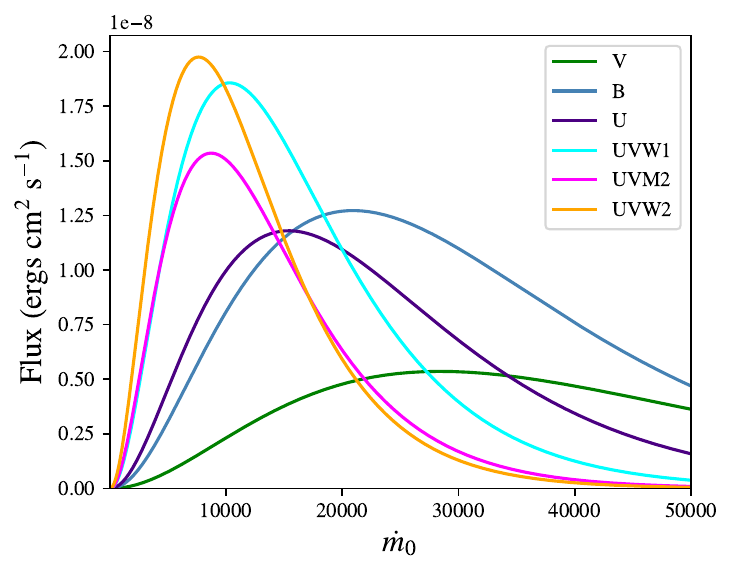}
    \caption[Mass accretion rate against flux in UVOT bands]{Flux in each of the UVOT bands as a
    function of the mass accretion rate $\dot{m}_0$. The peak flux is reached at $\dot{m}_0\sim 7000$ for the
    highest energy band (UVW2), much higher than is expected for ULXs; this means our UV view
    of high inclination ULXs, where the observed emission is from the wind photosphere,
    exists entirely within the region of UV emission on the left-hand side of the plot where the flux in a single band increases monotonically with $\dot{m}_0$.
    The plot was made using the \texttt{xspec} model
    \texttt{bbody} with a normalized luminosity and a temperature set to $T_{\mathrm{ph}}$ (eq \ref{eq:T_sph}),
    $m = 1.4 \ M_{\odot}$, $\beta = \zeta = 1$, $f_{\mathrm{col}}=1.7$ and 
    $\epsilon_{\mathrm{w}}=0.95$.}
    \label{fig:mdot_vs_flux}
\end{figure}

\noindent where $f_{\rm col}$ is a colour temperature correction factor,
$\zeta$, $\beta$ and $\epsilon_{\rm w}$ are constants relating to the wind-cone
opening angle, outflow velocity and fractional energy content in powering the
wind, $m$ is the accretor mass in M$_{\odot}$, and $\dot{m}_{0}$ is the
Eddington-scaled accretion rate.

By assuming the wind photosphere radiates as a blackbody at the characteristic
temperature $T_{\mathrm{sph}}$, that $\zeta\beta/\epsilon_{\rm w} \approx 1$ 
(A simplification that can be made as all parameters can be considered of order unity),
$f_{\mathrm{col}} \approx 2$ (see \citealt{1995_Shimura_ApJ...445..780S}), 
and accretor masses of 10 M$_{\odot}$ and 1.4 M$_{\odot}$, the UV emission in
the highest energy band (UVW2) (taking the form of a blackbody, peaking around 3 kT)
will increase in brightness until accretion rates in excess of $\dot{m}_{0}$ $\approx$
5000 $\times$ Eddington are reached for a 10 M$_{\odot}$ black hole and $\dot{m}_{0}$ $\approx$
7000 $\times$ Eddington M$_{\odot}$ for a 1.4 M$_{\odot}$ neutron star (see figure 
\ref{fig:mdot_vs_flux}). Such rates are safely above those inferred for known ULXs
\citep{2020_Vasilopoulos_MNRAS.491.4949V}.
Below these limits, and for an orientation
permitting a view into the wind-cone, one would expect a positive correlation
between the X-rays and UV. It is conceivable that an observer could be oriented
such that the closing of the wind cone inhibits their ability to see the
collimated emission. In this case, there would be a change in the ratio between
hard and soft X-ray emission (the soft being relatively more visible)
accompanying a drop in $T_{\rm sph}$. This would result in an anti-correlation
between $T_{\rm sph}$ and the UV brightness and a more complex correlation with
spectral hardness.  

For a fixed accretion rate, a change in the inclination of the disc/wind,
driven by precession (e.g. \citealt{2018_Middleton_MNRAS.475..154M,
2019_Middleton_MNRAS.489..282M}) would result in changes to the X-ray spectral
colours similar to the case of the wind-cone closing (see
\citealt{2015_Middleton_MNRAS.454.3134M}). In short, when the cone precesses
to higher inclinations to our line-of-sight, the X-ray emission should
diminish, and the low frequency emission should become brighter, leading to an
anti-correlation between the X-ray and UV emission (assuming the accretion rate
is high enough for the wind photosphere to emit substantially at such low
energies).

\subsection{Irradiated outer disc}

\begin{figure}
    \centering
    \includegraphics[width=\columnwidth]{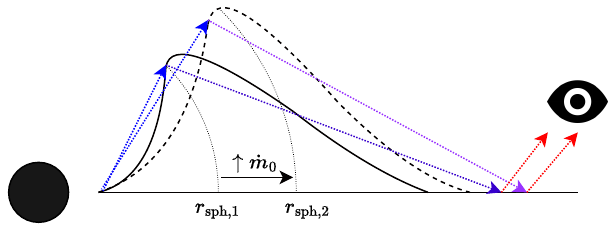}
    \caption[Irradiated Outer Disk]{Schematic for the geometry of an irradiated outer disc,
    an increase in the mass accretion rate results in the spherization radius
    moving outwards and decreasing in temperature. High energy photons
    produced in the inner parts of the accretion flow are 
     scattered by the large scale-height wind cone.
    }
    \label{fig:irradiated_disc}
\end{figure}

It has been suggested that the outer accretion disc could be irradiated by
X-ray emission from the inner regions after scattering by the wind
\citep{2013_Sutton_MNRAS.435.1758S}. As long as this irradiating SED has
sufficient intensity above 2~keV, down-scattering of these photons can produce
a UV-shoulder \citep{2008_Gierlinski_MNRAS.388..753G}. Exploring irradiation
requires RMHD simulations and radiative transfer calculations to follow the
photons from the inner regions to the outer disc.  However, such calculations
have yet to be performed, and so we instead base our reasoning on a simplified
picture (see figure \ref{fig:irradiated_disc}).

Should the accretor be a black hole or a low dipole field neutron star ($\lesssim
10^{9} G$), then the spectrum from the innermost regions is insensitive to
accretion rate \citep{2007_Poutanen_MNRAS.377.1187P} due to mass loss at larger
radii. If the ULX contains a high dipole field-strength neutron star (up to
$\sim$10$^{13}$G), the emergent spectrum above 2~keV is predicted to be
dominated by emission from the accretion column, with photons scattering to
escape the magnetosphere \citep{2017_Mushtukov_MNRAS.467.1202M}. Assuming the
magnetospheric radius lies within $r_{\rm sph}$, then the accretion rate
through the magnetosphere is mostly insensitive to the accretion rate at larger
radii (see e.g. \citealt{2020_King_MNRAS.494.3611K} although see also
\citealt{2019_Chashkina_A&A...626A..18C}).  Assuming the intrinsic spectrum
from the accretion column above 2~keV (see \citealt{2016_Brightman_M82_pulse,
2018_Walton_pulses}) does not change, we need only consider the changes in the
scattering medium between the outer disc and inner regions. As the accretion
rate at large radius increases, the wind cone closes and the optical depth of
the wind increases (see figure \ref{fig:irradiated_disc}).  There are then more
scatterings within the wind cone, which reduces the energy of photons created
in the inner regions; any escaping photons are therefore likely to be at lower
energy and less likely to thermalise in the outer disc
\citep{2008_Gierlinski_MNRAS.388..753G}. The converse is true for a drop in
accretion rate. 

For a fixed observer inclination but a varying accretion rate, the
presence (or lack of) a correlation between the X-rays and UV resulting from an
irradiated outer disc once again depends on whether an observer can see into
the wind cone. Should an observer be able to view the innermost regions
directly, then an increase in accretion rate will lead to an increase in X-ray
flux (and decrease in T$_{\rm sph}$) and a drop in UV flux, as fewer hard
X-rays impinge on the outer disc. Should an observer instead view the inner flow at higher
inclinations, then a more complex change in spectral hardness will result (as
described in \citealt{2015_Middleton_MNRAS.454.3134M}). For a fixed accretion
rate, a change in inclination of the inner disc and wind will not change the UV
emission (even if the wind were to tilt away from us, it would still irradiate
the far side of the disc) and the UV and X-rays will be uncorrelated. 

\begin{table}
\begin{tabular}{lllll}
\hline
                      &                         &                      & \multicolumn{2}{c}{\textbf{Flux}} \\
\textbf{Prediction}   & $\dot{m}$               & \textbf{Inclination} & \textbf{X-ray}    & \textbf{UV}   \\
\hline
Wind Photosphere      & Low                     & Low (face-on)        & \bluebar{20}      & \redbar{10}   \\
Wind Photosphere      & Low                     & Intermediate         & \bluebar{10}      & \redbar{20}   \\
Wind Photosphere      & Low                     & High (Edge-on)       & \bluebar{1}       & \redbar{20}   \\
                      &                         &                      &                   &               \\
Wind Photosphere      & High                    & Low (face-on)        & \bluebar{30}      & \redbar{20}   \\
Wind Photosphere      & High                    & Intermediate         & \bluebar{20}      & \redbar{20}   \\
Wind Photosphere      & High                    & High (Edge-on)       & \bluebar{1}       & \redbar{30}   \\
                      &                         &                      &                   &               \\
Irradiated outer disc & Low                     & Low (face-on)        & \bluebar{20}      & \redbar{20}   \\
Irradiated outer disc & Low                     & Intermediate         & \bluebar{10}      & \redbar{20}   \\
Irradiated outer disc & Low                     & High (Edge-on)       & \bluebar{1}       & \redbar{20}   \\
                      &                         &                      &                   &               \\
Irradiated outer disc & High                    & Low (face-on)        & \bluebar{30}      & \redbar{10}   \\
Irradiated outer disc & High                    & Intermediate         & \bluebar{20}      & \redbar{10}   \\
Irradiated outer disc & High                    & High (Edge-on)       & \bluebar{1}       & \redbar{10}   \\  
\hline
\end{tabular}
\label{tab:predictions}
\caption{Predictions of the relative X-ray to UV emission originating from the wind photosphere and irradiation of the outer disc. The blue and red bars provide a visual representation of the relative observed intensity of the
emitted radiation in the X-ray and UV bands. These estimates are approximate and quantized to four discrete lengths to serve as a qualitative indicator.}
\end{table}

\subsection{Irradiated companion star}

%\begin{figure}
%    \centering
%    \includegraphics[width=\columnwidth]{figures/irr_area_theta.pdf}
%    \caption{}
%    \label{fig:irr_area_theta}
%\end{figure}

\begin{figure}
    \centering
    \includegraphics[width=\columnwidth]{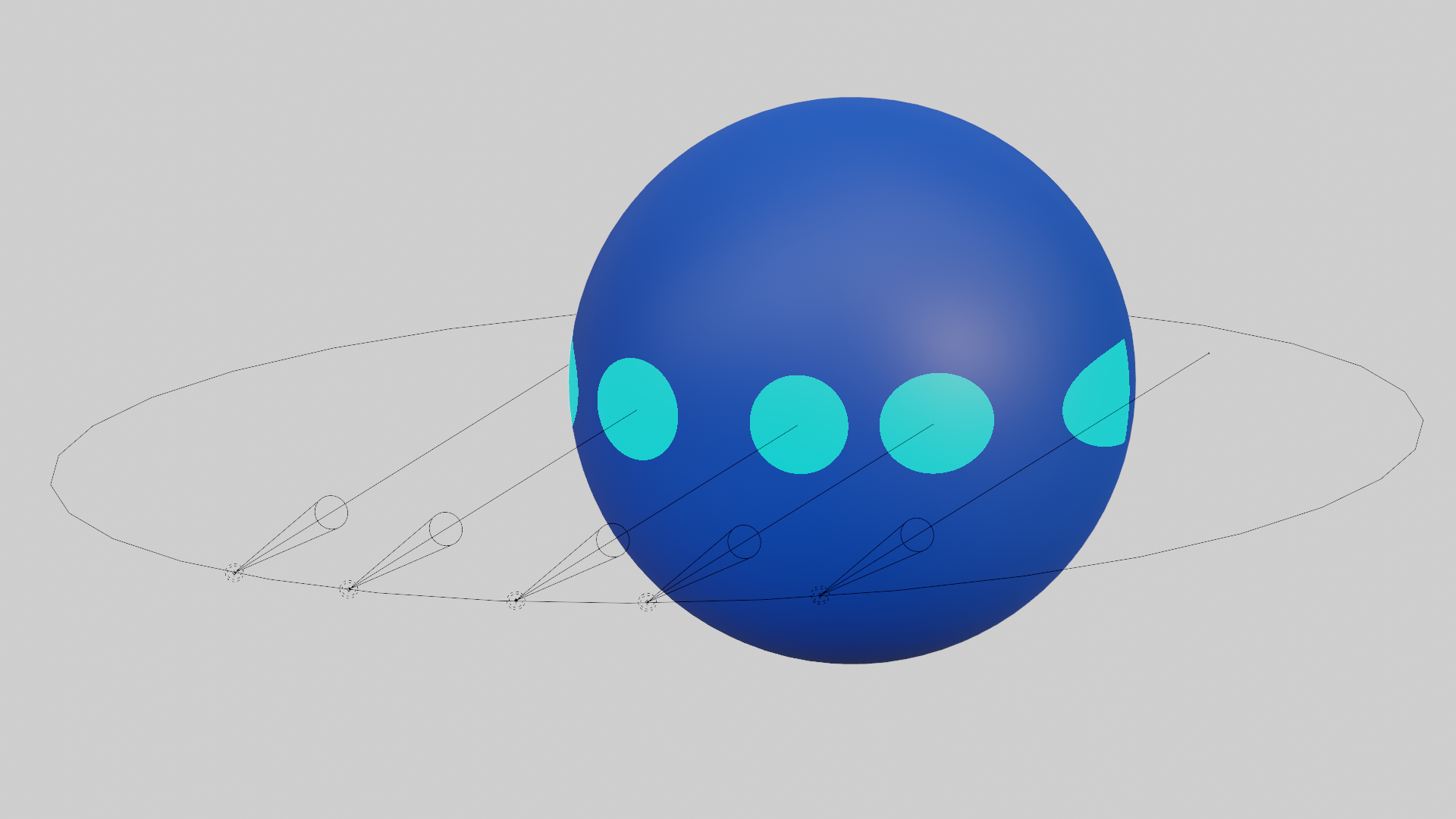}
    \caption{Schematic of irradiation of the secondary star as a tight cone of X-ray emission
    travelling on an elliptical orbit.}
    \label{fig:irr_star_orbit}
\end{figure}

A third option, distinct from the above two -- and proposed to explain the
anti-phase optical super-orbital period seen in NGC 7793 P13
\citep{2021_Furst_A&A...651A..75F} -- is that the X-ray cone sweeps over the
companion star and hard X-rays thermalise in the outer layers of the star leading to
enhanced low frequency emission \citep{2014_Motch_Natur.514..198M}. To explore this scenario,
we simulate a cone of X-ray emission irradiating a star at different orbital phases and
different orientations relative to each other and the observer.
Figure \ref{fig:irr_star_orbit} shows a schematic of how the irradiated area
(shown in light blue) from a cone of X-ray emission may vary as it sweeps over the star.
The irradiated projected area seen by an observer is subject to many variables, including
the size and shape of the orbit, star and cone, as well as the position of the observer.
For further elaboration on the specific implementations of the irradiated companion model,
please refer to Appendix \ref{sec:irradiated_companion}.

\begin{figure}
    \centering
    \includegraphics[width=\columnwidth]{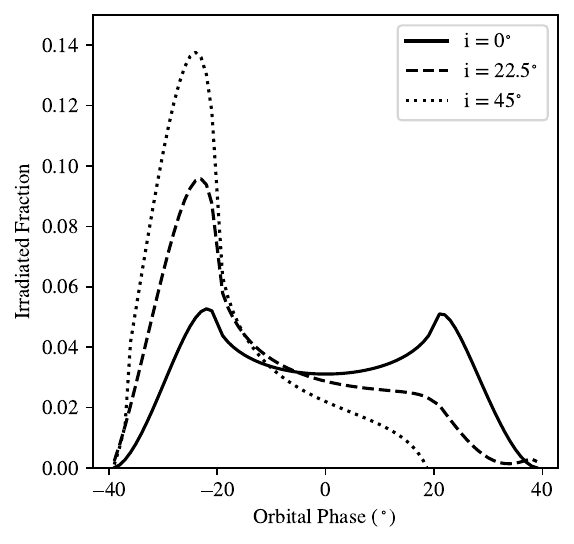}
    \caption{How the irradiated fraction, defined as $A_{\mathrm{proj}} / \pi R_{\star}^{2}$, varies
    as a function of orbital phase transiting through
    a cone of emission with opening angle $\theta = 20^{\circ}$. The transit is shown
    for three different observer inclinations, $i = 0, 22.5$ and $45$ degrees.
    A ratio of $R_{\star}/a = 0.5$ is assumed.
    }
    \label{fig:irr_area_incl}
\end{figure}

Figure \ref{fig:irr_area_incl} shows how the irradiated fraction, defined as
the projected area seen by the observer divided by the maximum possible projected
area $A_{\mathrm{proj}}/\pi R_{\star}^{2}$, varies as a function of the
(circular) orbital phase. The transit is simulated for a cone sweeping directly
through the centre of the star and is shown for three different observer
inclinations. It is worth noting that the inclination parameter alone is not sufficient to
fully describe the configuration of the system and the resulting irradiated area, this is
as the entire system can be rotated while the inclination remains unchanged 
resulting in different projected areas for the same inclination value. 
Never the less, for illustrative purposes figure \ref{fig:irr_area_incl} shows how the profile of a
transit may be significantly altered by the observer's position.

If we were to assume that the axis of the wind-cone is perpendicular to the
orbital axis, only wind-cone half-opening angles ($\theta/2$) which satisfy
$\theta/2 > \mathrm{arccos}(R_{\star}/a)$ would be able to irradiate the star (where
$R_{\star}$ is the radius of the companion star and $a$ is the average
semi-major axis of the orbit). For example, a half-opening angle of $10^{\circ}$
would require a $R_{\star}/a \approx $ 1 which poses issues for stable accretion.
Allowing for larger opening angles will of course lower the value of the ratio of
$R_{\star}/a$ although, as super-critical accretion demands a half-opening angle of
at least $45^{\circ}$ (where the scale-height of the disc is approximately unity),
this requires $R_{\star}/a >$ 0.7. Thus, large amounts of irradiation likely require
an offset between the axis of the wind cone and that of the orbit (which may naturally lead to
precession of the wind cone itself \citealt{2018_Middleton_MNRAS.475..154M,
2019_Middleton_MNRAS.489..282M}). 

Estimates for the eccentricity of ULX orbits have been obtained for a few pulsating
sources, where the pulse arrival time allows for the modelling of the orbital
parameters of the system. Based on modelling of the X-ray pulsations in NGC
7793 P13, \citep{2018_Furst_A&A...616A.186F} find an eccentricity of $e \le
0.15$, whilst \citep{2014_Bachetti_Natur.514..202B} found that M82 X-2 had
a near circular orbit of $e = 0.003$. In the absence of any detailed orbital
information, we therefore assume a simple circular orbit in our model.

We have investigated the potential impact of an irradiated companion, illuminated by a cone
of X-ray emission, on the observed UV/optical emission in ULXs. Our model is subject to limitations
and uncertainties, and it is possible that specific conditions may be required for its geometry to
be a reality in these sources. However, despite these challenges, it may be possible in the future
to determine the ephemeris based on the observed profiles seen in the UV/optical light curves of
these sources. High-resolution spectroscopy can also be used to determine whether significant heating
of the star's outer layers is present from specific spectral lines. In the next section, we will
describe our approach to obtaining observations and reducing data to create long-term light curves
for a sample of ULXs.

\section{Observations and Data Reduction}

\begin{table*}
\begin{tabular}{lcccccccc}
\hline
Source Name        & RA            & DEC           & $\lambda$ & Position reference                         & $D$      & $D$ meth. & $D$ reference                                  & UV \\
                   & "h:m:s"       & "d:m:s"       &           &                                 & $\mathrm{Mpc}$ &    &                                          & Emis.         \\
\hline
V404 Cygni         & 20 24 03.8254 & +33 52 01.962 & O   & \cite{2018_Gaia_yCat.1345....0G}      & 0.0023   & parallax & \cite{2009_Miller_ApJ...706L.230M}       & PL\\
Swift J0243.6+6124 & 02 43 40.4252 & +61 26 03.757 & O   & \cite{2020_Gaia_yCat.1350....0G}      & 0.0055   & parallax & \cite{2020_Gaia_yCat.1350....0G}         & PL\\
SS433              & 19 11 49.5647 & +04 58 57.827 & O   & \cite{2020_Gaia_yCat.1350....0G}      & 0.0055   &          & \cite{2004_Blundell_ApJ...616L.159B}     & PL\\
SMC X-3            & 00 52 05.6251 & -72 26 04.228 & O   & \cite{2020_Gaia_yCat.1350....0G}      & 0.0600   & cepheid  & \cite{2017_Karachentsev_AJ....153....6K} & PL\\
IC10 X-1           & 00 20 29.09   & +59 16 51.9   & X   & \cite{2004_Bauer_ApJ...601L..67B}     & 0.7943   & redshift & \cite{2019_Lianou_AA...631A..38L}        & N\\
M31 ULX-1          & 00 42 53.15   & +41 14 22.9   & X   & \cite{2012_Kaur_AA...538A..49K}       & 0.8200   & T-RDB    & \cite{2017_Karachentsev_AJ....153....6K} & N\\
M33 ULX-1          & 01 33 50.8965 & +30 39 36.630 & O   & \cite{2020_Gaia_yCat.1350....0G}      & 0.9300   & T-RDB    & \cite{2017_Karachentsev_AJ....153....6K} & PL\\
NGC300 ULX-1       & 00 55 04.86   & -37 41 43.7   & O   & \cite{2008_Barbon_yCat....1.2024B}    & 2.0230   & redshift & \cite{2019_Lianou_AA...631A..38L}        & PL\\
NGC55 ULX          & 00 15 28.89   & -39 13 18.8   & X   & \cite{2012_Lin_ApJ...756...27L}       & 2.1100   & T-RDB    & \cite{2017_Karachentsev_AJ....153....6K} & Edg\\
IC342 ULX-1        & 03 45 55.612  & +68 04 55.29  & O   & \cite{2008_Feng_ApJ...675.1067F}      & 3.4356   & redshift & \cite{2019_Lianou_AA...631A..38L}        & N\\
IC342 ULX-2        & 03 46 15.61   & +68 11 12.8   & X   & \cite{2014_Heida_MNRAS.442.1054H}     & 3.4356   & redshift & \cite{2019_Lianou_AA...631A..38L}        & N\\
NGC4945 XMM-1      & 13 05 32.89   & -49 27 34.1   & X   & \cite{2004_Swartz_ApJS..154..519S}    & 3.4674   & redshift & \cite{2019_Lianou_AA...631A..38L}        & E\\
Holmberg II X-1    & 08 19 28.99   & +70 42 19.4   & X   & \cite{2014_Heida_MNRAS.442.1054H}     & 3.4674   & redshift & \cite{2019_Lianou_AA...631A..38L}        & E\\
M81 X-6            & 09 55 32.95   & +69 00 33.6   & X   & \cite{2014_Heida_MNRAS.442.1054H}     & 3.5975   & redshift & \cite{2019_Lianou_AA...631A..38L}        & E\\
M82 X-2            & 09 55 51.040  & +69 40 45.49  & X   & \cite{2006_Kaaret_ApJ...646..174K}    & 3.6141   & redshift & \cite{2019_Lianou_AA...631A..38L}        & E\\
NGC253 X-2         & 00 47 32.97   & -25 17 50.2   & X   & \cite{2005_Liu_ApJS..157...59L}       & 3.6983   & redshift & \cite{2019_Lianou_AA...631A..38L}        & E\\
NGC253 X-9         & 00 47 22.59   & -25 20 50.9   & X   & \cite{2014_Heida_MNRAS.442.1054H}     & 3.6983   & redshift & \cite{2019_Lianou_AA...631A..38L}        & N\\
NGC247 ULX-1       & 00 47 04.00   & -20 47 45.7   & X   & \cite{2005_Liu_ApJS..157...59L}       & 3.7200   & T-RDB    & \cite{2017_Karachentsev_AJ....153....6K} & N\\
NGC7793 P13        & 23 57 50.90   & -32 37 26.6   & X   & \cite{2011_Pannuti_AJ....142...20P}   & 3.7325   & redshift & \cite{2019_Lianou_AA...631A..38L}        & PL\\
Holmberg IX X-1    & 09 57 53.290  & +69 03 48.20  & O   & \cite{2009_Abazajian_ApJS..182..543A} & 3.8500   & T-RDB    & \cite{2017_Karachentsev_AJ....153....6K} & E\\
NGC1313 X-1        & 03 18 20.00   & -66 29 10.9   & X   & \cite{2014_Heida_MNRAS.442.1054H}     & 4.2500   &          & \cite{2016_Tully_AJ....152...50T}        & N\\
NGC1313 X-2        & 03 18 22.00   & -66 36 04.3   & X   & \cite{2005_Liu_ApJS..157...59L}       & 4.2500   &          & \cite{2016_Tully_AJ....152...50T}        & E\\
NGC5204 ULX-1      & 13 29 38.62   & +58 25 05.6   & X   & \cite{2014_Heida_MNRAS.442.1054H}     & 4.5900   & T-RDB    & \cite{2017_Karachentsev_AJ....153....6K} & E\\
UGC6456 ULX        & 11 28 03.000  & +78 59 53.41  & O   & \cite{2020_Vinokurov_ApJ...893L..28V} & 4.6300   & T-RDB    & \cite{2017_Karachentsev_AJ....153....6K} & E\\
NGC4395 ULX-1      & 12 26 01.53   & +33 31 30.6   & X   & \cite{2014_Heida_MNRAS.442.1054H}     & 4.7600   &          & \cite{2016_Tully_AJ....152...50T}        & E\\
M83 ULX-1          & 13 37 05.13   & -29 52 07.1   & X   & \cite{2014_Long_ApJS..212...21L}      & 4.8978   & redshift & \cite{2019_Lianou_AA...631A..38L}        & N\\
M83 ULX-2          & 13 37 20.12   & -29 53 47.7   & X   & \cite{2005_Liu_ApJS..157...59L}       & 4.8978   & redshift & \cite{2019_Lianou_AA...631A..38L}        & E\\
NGC5408 ULX-1      & 14 03 19.63   & -41 22 58.7   & X   & \cite{2014_Heida_MNRAS.442.1054H}     & 5.3211   & redshift & \cite{2019_Lianou_AA...631A..38L}        & N\\
NGC6946 ULX-1      & 20 35 00.11   & +60 09 08.5   & X   & \cite{2012_Lin_ApJ...756...27L}       & 6.7298   & redshift & \cite{2019_Lianou_AA...631A..38L}        & E\\
NGC6946 ULX-3      & 20 35 00.74   & +60 11 30.6   & X   & \cite{2004_Swartz_ApJS..154..519S}    & 6.7298   & redshift & \cite{2019_Lianou_AA...631A..38L}        & PL\\
M101 ULX-1         & 14 03 32.38   & +54 21 03.0   & X   & \cite{2014_Heida_MNRAS.442.1054H}     & 7.1121   & redshift & \cite{2019_Lianou_AA...631A..38L}        & N\\
NGC4559 ULX-1      & 12 35 51.71   & +27 56 04.1   & X   & \cite{2014_Heida_MNRAS.442.1054H}     & 7.1450   & redshift & \cite{2019_Lianou_AA...631A..38L}        & E\\
M51 ULX-7          & 13 30 01.01   & +47 13 43.9   & X   & \cite{2014_Heida_MNRAS.442.1054H}     & 7.6000   & redshift & \cite{2011_Cappellari_MNRAS.413..813C}   & E\\
NGC5585 ULX        & 14 19 39.39   & +56 41 37.8   & X   & \cite{2014_Heida_MNRAS.442.1054H}     & 7.8300   &          & \cite{2016_Tully_AJ....152...50T}        & PL\\
NGC925 ULX-1       & 02 27 27.53   & +33 34 42.9   & X   & \cite{2014_Heida_MNRAS.442.1054H}     & 9.2045   & redshift & \cite{2019_Lianou_AA...631A..38L}        & E\\
NGC925 ULX-2       & 02 27 21.52   & +33 35 00.8   & X   & \cite{2014_Heida_MNRAS.442.1054H}     & 9.2045   & redshift & \cite{2019_Lianou_AA...631A..38L}        & E\\
NGC7090 ULX-3      & 21 36 31.94   & -54 33 57.2   & X   & \cite{2012_Lin_ApJ...756...27L}       & 9.5060   & redshift & \cite{2019_Lianou_AA...631A..38L}        & Edg\\
NGC5907 ULX        & 15 15 58.60   & +56 18 10.0   & X   & \cite{2011_Swartz_ApJ...741...49S}    & 17.2187  & redshift & \cite{2019_Lianou_AA...631A..38L}        & Edg\\
NGC1365 X-1        & 03 33 34.60   & -36 09 35.0   & X   & \cite{2005_Liu_ApJS..157...59L}       & 17.2982  & redshift & \cite{2019_Lianou_AA...631A..38L}        & N\\
NGC1365 X-2        & 03 33 41.85   & -36 07 31.4   & X/O & \cite{2009_Strateva_ApJ...692..443S}  & 17.2982  & redshift & \cite{2019_Lianou_AA...631A..38L}        & PL\\
NGC1042 ULX-1      & 02 40 25.62   & -08 24 28.9   & X   & \cite{2012_Lin_ApJ...756...27L}       & 19.2000  &          & \cite{2007_Oey_ApJ...661..801O}          & N\\
ESO 243-49 HLX-1   & 01 10 28.30   & -46 04 22.3   & X   & \cite{2010_Webb_ApJ...712L.107W}      & 115.3500 &          & \cite{2016_Tully_AJ....152...50T}        & N\\
\hline
\end{tabular}
\caption{Source information for those ULXs investigated in this paper, sorted by host galaxy 
distance. $\lambda$ corresponds to the wavelength (X-ray or optical) that the position was acquired in.
The `D meth.' column denotes the method used for determining the distance (where T-RDB is
Tip of the red-giant branch). The `UV Emis.' column corresponds to the spatial extent of the
UV/optical emission based on visual inspection of the stacked UVOT images.
}
\label{tab:source_list}
\end{table*}

The Neil Gehrels Swift Observatory (\textit{Swift})
\citep{2004_Gehrels_ApJ...611.1005G} is capable of observing simultaneously in
both the UV and X-ray bands via its UVOT \citep{2005_Roming_SSRv..120...95R}
and XRT \citep{2005_Burrows_SSRv..120..165B} cameras, and so provides a unique
opportunity to explore whether observations of ULXs are consistent with any of the
predictions discussed above. We begin by creating our ULX sample by
crossmatching several ULX catalogues, \citep{2019_Earnshaw_MNRAS.483.5554E,
2020_Kovlakas_MNRAS.498.4790K, 2022_Bernadich_AA...659A.188B,
2022_Walton_MNRAS.509.1587W} with the \textit{Swift} Master Catalogue
(\texttt{SWIFTMASTR}), accessible via {\sc
HEASARC}\footnote{\href{https://heasarc.gsfc.nasa.gov}{https://heasarc.gsfc.nasa.gov}}.
For the sake of comparison to nearby sources where data quality is high, we
also include three extensively studied Galactic sources which reach super-Eddington rates of accretion, Swift J0243.6+6124,
SS433 and V404 Cygni.  Swift J0243.6+6124 is known to contain a magnetised
neutron star and appear as a ULX \citep{2020_Eijnden_MNRAS.496.4127V}, SS433 is
widely considered to be an edge-on ULX (\citealt{2004_Fabrika_ASPRv..12....1F};
\citealt{2021_Middleton_MNRAS.506.1045M}) and V404 Cygni is a LMXB which
reached around or just in excess of its Eddington luminosity during its 2015
outburst (\citealt{2017_Motta_MNRAS.471.1797M}).

We locate all observations where the source lies within the nominal (23.6') XRT
field-of-view. Due to the differences between the XRT and UVOT field-of-view,
there is a mismatch between the number of observations in both bands.  We set
the requirement that there be 20 observations in both bands for a source to
appear in our sample.  For each source, we manually cross-matched with SIMBAD
to obtain distances and positions.  The final sample with  relevant
information is shown in Table \ref{tab:source_list}.

\subsection{XRT Data Reduction}

XRT data was extracted using the standard \textit{Swift}/XRT processing
pipeline \citep{2009_Evans_MNRAS.397.1177E}, using the SIMBAD coordinates of
the source, and the `simple' centroid method with a positional error of 1" (see 
\citealt{2009_Evans_MNRAS.397.1177E}).  We extract light curves in three bands,
\texttt{full} $(0.3 - 10.0 \ \mathrm{keV})$, \texttt{soft} $(0.3 - 1.5 \mathrm{keV})$ and \texttt{hard}
$(1.5 - 10.0 \mathrm{keV})$, with the time binning set at a single point per
observation. Using the \texttt{full} band data, we set the minimum source
significance -- defined as the number of counts in the source region divided by
the square root of the counts in the source and background ($\mathrm{C_{src}} /
\sqrt{\mathrm{C_{src}} + \mathrm{C_{bkg}}}$) -- to the default value of 3; observations where the
source is not detected at or above this threshold yield upper limits. The
pipeline additionally calculates the hardness ratio, defined as the ratio of
hard to soft count rates $HR = \mathrm{C_{hard}} / \mathrm{C_{soft}}$.
$1\sigma$ errors are provided for each of the above measurements.  %We also
%extract summed spectra by combining all the observations for each source using
%the %\textit{Swift} XRT pipeline.  
The pipeline also includes \texttt{GOOD} and
\texttt{BAD} values, which indicates whether the pipeline was able to obtain a
centroid in a given snapshot, (such that \texttt{BAD} values are potentially
unreliable).

\begin{figure*}
    \centering
    \includegraphics[width=0.8\textwidth]{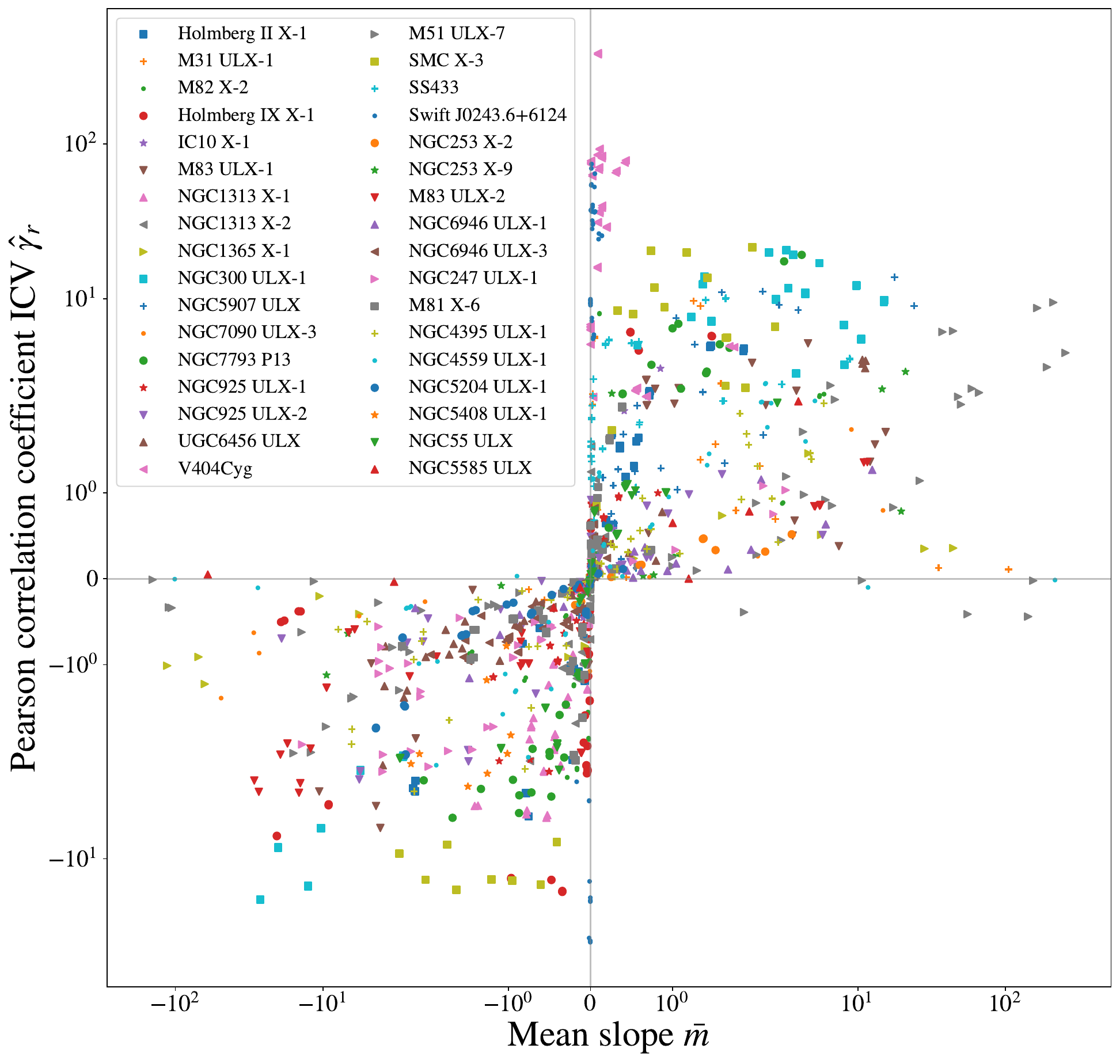}
    \caption[Mean Slope $\Bar{m}$ plotted against inverse coefficient of variation correlation coefficient
    $\hat{\gamma}_{r}$]{Mean slope $\Bar{m}$ (x) plotted against the inverse coefficient of variation of the Pearson
    correlation coefficient $\hat{\gamma}_{r}$ for all simulations.}
    \label{fig:mean_all_corr}
\end{figure*}

\subsection{UVOT Data Reduction}

UVOT data were processed locally using \texttt{HEASARC v6.29} and the November
8, 2021 CALDB.  We stacked all UVOT and XRT observations, combining them into a
single image and manually inspected to identify any clear counterparts.

For each source, a circular extraction region with a radius of 5" was centred
on the Swift XRT position, while background regions were manually positioned in
a contaminant-free location with a size of 15" as recommended by the UVOT data
analysis manual\footnote{
\href{https://swift.gsfc.nasa.gov/analysis/UVOT_swguide_v2_2.pdf}{https://swift.gsfc.nasa.gov/analysis/UVOT\_swguide\_v2\_2.pdf}}.
The distances between the centres of the UVOT and XRT (SIMBAD) source regions
as well as the position of the UVOT background regions are provided in the
supplementary material.

Level 2 UVOT images were processed locally using the \texttt{uvotimsum} to
combine all snapshot extensions, then \texttt{uvotsource} with a
signal-to-noise threshold of 3 was used to obtain photometric magnitudes in a
given observation. We then determined whether the source was detected using the
\texttt{NSIGMA} column provided as output from \texttt{uvotsource}.  All
observations of a given source were combined to produce a long term light-curve
in all of the UVOT bands: V (5469 {\AA}), B (4392 {\AA}), U (3465 {\AA}), UVW1
(2600 {\AA}), UVM2 (2246 {\AA}), UVW2 (1928 {\AA}).

\section{Analysis and Results}

\subsection{UV Counterparts}

The`UV Emis.' column in Table \ref{tab:source_list} indicates the
spatial nature of the UV emission of the source based upon the stacked images.
Sources with point-like UV counterparts and emission regions comparable to
\textit{Swift's} PSF are labelled as `PL', sources in or near regions of
extended UV emission as `E', those that do not display strong UV emission in
the images are labelled as `N', and three in edge-on galaxies `Edg' where point-sources
in the UV/optical are difficult to disentangle from the galactic emission.

\subsection{Testing for linear X-ray/UV correlations}

\begin{figure*}
    \centering
    \includegraphics{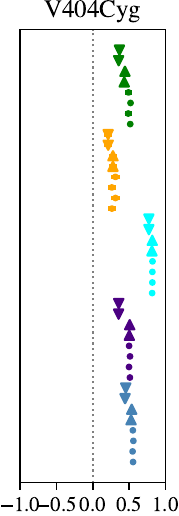}
    \includegraphics{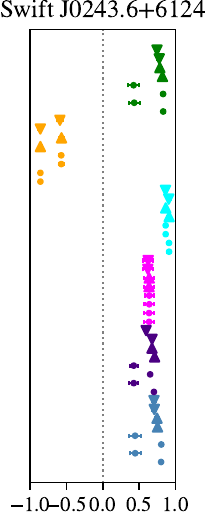}
    \includegraphics{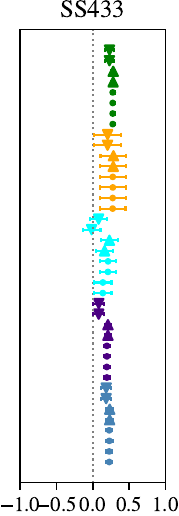}
    \includegraphics{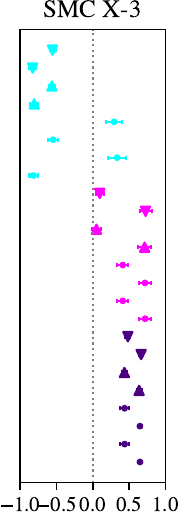}
    \includegraphics{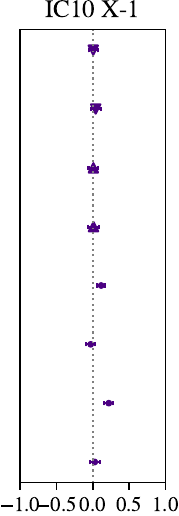}
    \includegraphics{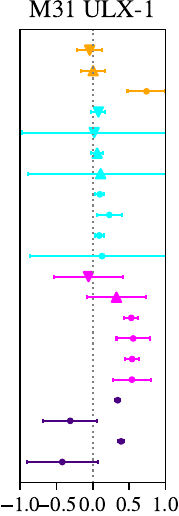}
    \includegraphics{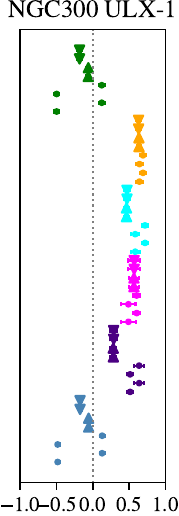}
    \includegraphics{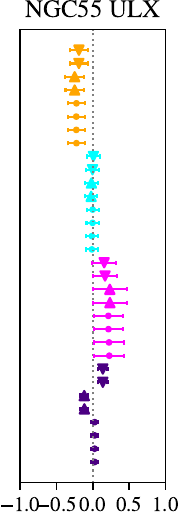}
    \includegraphics{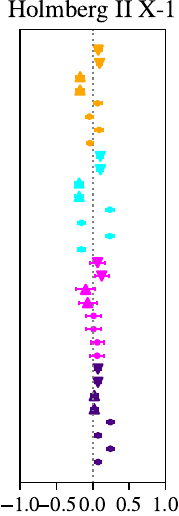}
    \includegraphics{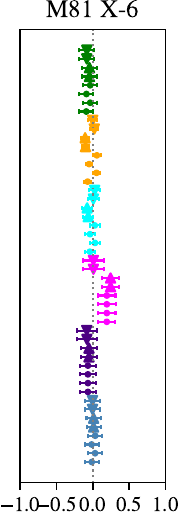}
    \caption{Distribution of the correlation coefficient for
             10 of the sources investigated.
             Error bars correspond to 1$\sigma$,
             calculated from 10,000 Monte Carlo iterations. Colours correspond to the different
             UVOT filters used (See legend in figures \ref{fig:blackbody_uvot} \& \ref{fig:mdot_vs_flux}) and are displayed in decreasing energy from top to bottom. The marker on the datapoint
             denotes which of the three XRT band with which the correlation was carried out with: 
             The full band is denoted with a dot ($\bullet$), the hard band by an upwards triangle ($\blacktriangle$) and the soft band with a downwards triangle ($\blacktriangledown$),
             the correlations against the hardness ratio are not shown here.
             The same colour and marker may appear multiple times
             as separate simulations were conducted with the 
             inclusion or exclusion of upper limits and/or bad data points.}
    \label{fig:correlations_1}
\end{figure*}

\begin{figure*}
    \centering
\includegraphics{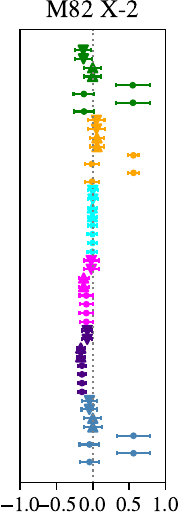}
\includegraphics{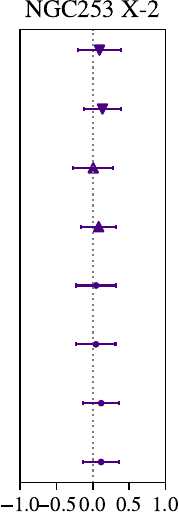}
\includegraphics{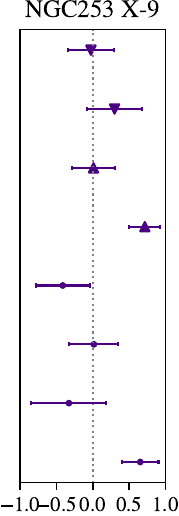}
\includegraphics{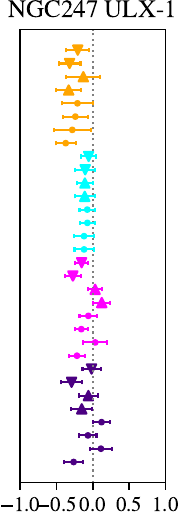}
\includegraphics{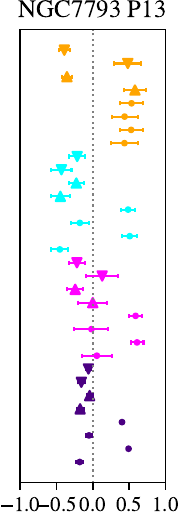}
\includegraphics{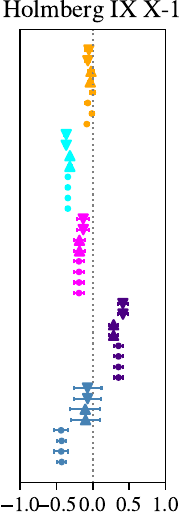}
\includegraphics{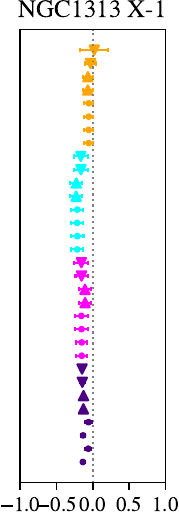}
\includegraphics{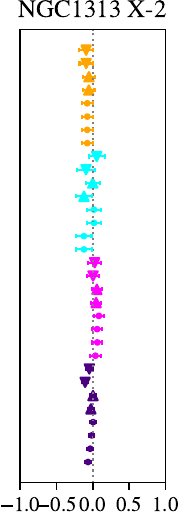}
\includegraphics{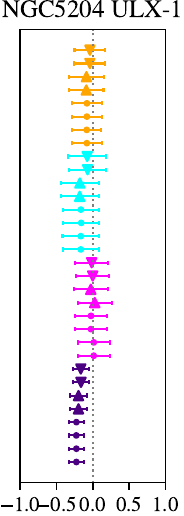}
\includegraphics{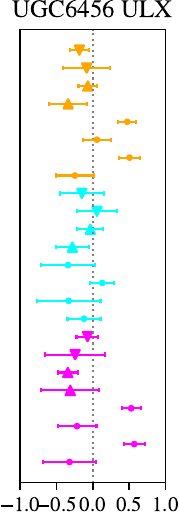}
    \caption{See caption for figure \ref{fig:correlations_1}.}
    \label{fig:correlations_2}
\end{figure*}
\begin{figure*}
    \centering
\includegraphics{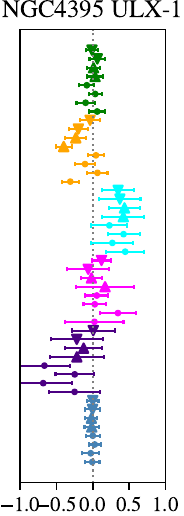}
\includegraphics{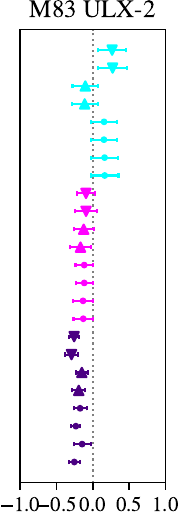}
\includegraphics{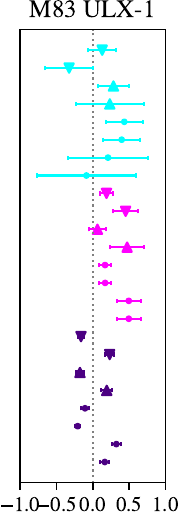}
\includegraphics{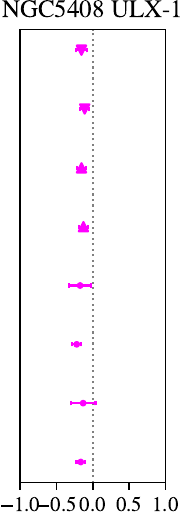}
\includegraphics{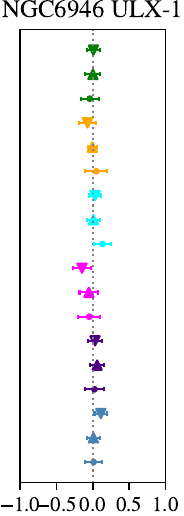}
\includegraphics{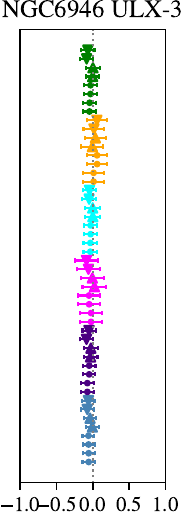}
\includegraphics{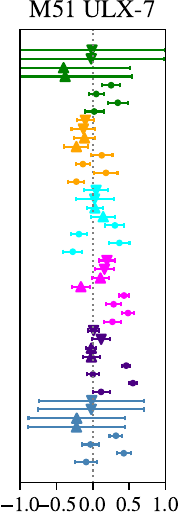}
\includegraphics{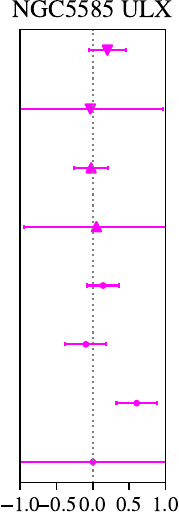}
\includegraphics{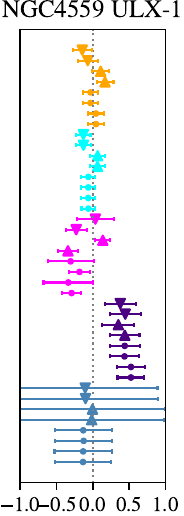}
\includegraphics{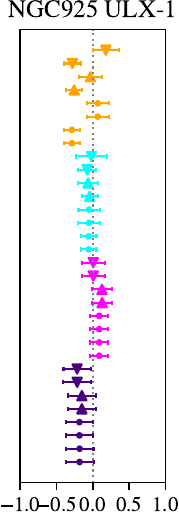}
    \caption{See caption for figure \ref{fig:correlations_1}.}
    \label{fig:correlations_3}
\end{figure*}
\begin{figure*}
    \centering
\includegraphics{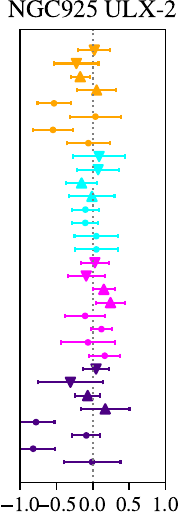}
\includegraphics{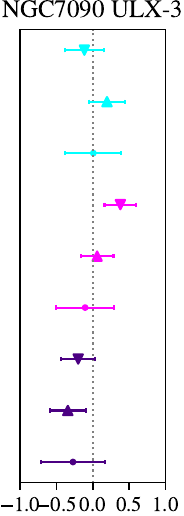}
\includegraphics{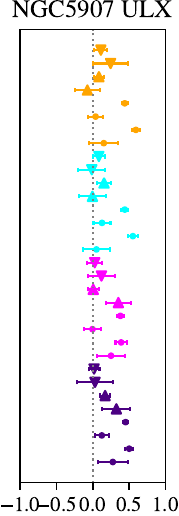}
\includegraphics{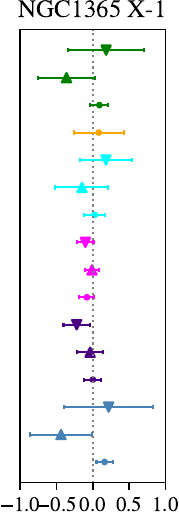}
    \caption{See caption for figure \ref{fig:correlations_1}.}
    \label{fig:correlations_4}
\end{figure*}

A cut of $\pm 5 \sigma$ around the mean count rate was
applied to both the UVOT and XRT data prior to simulating. This cut resulted in only
5 observations being excluded due to UVOT outliers, which were due to observations
where the source was close to the edge of the detector or the background region was located
within the fits image but outside the detector plate. Around $\sim 30$ observations were
excluded due to XRT outliers, some of these are instrumental however others may be real and
be due to source flaring, we find that around 0-3 observations are excluded for each lightcurve 
that may be composed of several hundred data points, the removal of these outliers results in marginally
weaker detected correlations meaning that our results consistute a slightly more conservative estimate.
To search for linear correlations in the clean X-ray and UVOT light curves, we employ the following
method: for each observation data point we sampled the $1\sigma$ error on the
count rate in both the XRT and UVOT bands assuming Gaussian distributions; if
the observation contained an upper limit, the data point was sampled assuming a
uniform distribution between 0 and this value. This sampling allows us to
obtain realisations of the light curve with the same time-sampling as the
original. We then performed a least-squares fit with a straight line of the
form $y=mx+c$ to the re-sampled data points and calculated the Pearson
correlation coefficient $r$ given by:

\begin{equation}
    r = \frac{\sum_{i=1}^{n} (x_i - \Bar{x})(y_i - \Bar{y})}{\sqrt{\sum_{i=1}^{n} (x_i - \Bar{x})^2 }\sqrt{\sum_{i=1}^{n} (y_i - \Bar{y})^2}}
    \label{eq:pearson}
\end{equation}
\noindent where $x_{i}$ and $y_{i}$ are the ith values in the sample, while
$\Bar{x}$ and $\Bar{y}$ are the means over all the $n$ data points. This
resampling and fitting process was repeated 10,000 times to obtain posterior
distributions for $r$, $m$ and $c$, from which the
mean and standard deviation were calculated.

To assess how well constrained are the posterior distributions for
the fit parameters, we calculate the inverse coefficient of variation (ICV)
($\hat{\gamma}_{\mathrm{par}}$) by dividing the mean of the fit parameters
($\overline{\mathrm{par}}$) by their standard deviation
($\sigma_{\mathrm{par}}$), i.e. $\hat{\gamma}_{\mathrm{par}} =
\overline{\mathrm{par}} / \sigma_{\mathrm{par}}$.  The absolute value of
$\hat{\gamma}_{\mathrm{par}}$ may be interpreted as a significance value, with
higher values corresponding to more strongly peaked distributions (i.e. the
distribution has less scatter).

Simulations were carried out on all possible combinations of the four X-ray bands
(\texttt{full}, \texttt{soft}, \texttt{hard}, \texttt{HR}), and the six UVOT
filters. Simulations were additionally carried out including and excluding
observations labelled as \texttt{BAD} in the XRT pipeline, as well as data
points labelled as upper limits in the \texttt{full} XRT band. The simulation
grid implies that a single source may have a total of 24 possible simulation
combinations for the \texttt{full} XRT band, and 36 for the remaining XRT bands
(\texttt{hard}, \texttt{soft} and \texttt{HR}), giving a maximum of 60
correlations per source, assuming that the source has been visited
in all UVOT bands and its XRT light curves contain both \texttt{BAD} and upper
limit data points. In practice this is not the case and the total number of
sets of simulations for each source is almost always lower than 60.

Figure \ref{fig:mean_all_corr} shows the mean value of the slope ($m$) and the
ICV of the Pearson correlation coefficient ($\hat{\gamma}_{r}$) across all of our
simulations. Sources appear multiple times on the plot due to the
aforementioned grid in simulation space.  The diagonal distance from the centre in this
parameter space might provide an indication that a correlation exists -- the
upper right corresponding to positive correlations, the lower left to negative.
However as we will demonstrate, figure \ref{fig:mean_all_corr} may not encode
sufficient information to locate more physical correlations.

Figures \ref{fig:correlations_1}, \ref{fig:correlations_2}, \ref{fig:correlations_3}
and \ref{fig:correlations_4} show the correlation coefficient obtained from our simulations
for different X-ray bands and UVOT filters. Each errorbar corresponds to a discrete simulation
run and is coloured based on the UVOT filter used in the correlation and has a marker corresponding
to the XRT band it was correlated against (see caption in figure \ref{fig:correlations_1}).
If the errorbars for a single source consistently deviate far from the dotted line (zero) and are well constrained,
then it indicates a consistent correlation between the UVOT filter and the X-ray bands.
The sources V404 Cygni, Swift J0243.6+6124, SS433, SMC X-3, NGC300 ULX-1, NGC 7793 P13 and Holmberg IX X-1
show the clearest cases of such correlation. Some errorbars, such as those seen for
the V band in M51 ULX-7, have large sizes due to non-Gaussian distributions for $r$ over the
10,000 Monte-Carlo simulations.

\section{Discussion \& Conclusions} \label{sec:discussion}

Although ULXs are defined empirically by their X-ray luminosity, they are
well-known to emit over a broad energy range. Indeed, bright optical/UV
emission (in excess of 10$^{39} \ \ergss$) is observed to originate in both
Galactic super-critical accretors (SS433: \citealt{1997_Dolan_A&A...327..648D})
and well studied ULXs (NGC 6946 ULX-3: \citealt{2010_Kaaret_ApJ...714L.167K})
with a mixture of potential origins. In this paper, we have made predictions
about how the UV and X-ray emission might correlate (or anti-correlate) depending
on the dominant mechanism for the low frequency radiation: the irradiated donor star,
irradiated outer disc or wind photosphere. 

Based on simple arguments, we predict a lack of any correlation between the UV
and X-ray emission where the star is irradiated by a
wind-cone, as, with the exception of an orientation where the central regions
are eclipsed, the X-ray emission should remain constant (in the absence of
precession or mass accretion rate changes). In the case of disc irradiation or
precession of the super-critical disc/wind, the exact nature of the correlation
depends mostly on the observer inclination and any changes in accretion rate at
large radius. Certainly for a fixed inclination angle (again, in the absence of
precession), a negative correlation would be expected for disc irradiation as
the X-ray emission (assumed here to originate within the wind-cone) increases
with increasing $\dot{m}$, but the optical depth to the outer regions also
increases. A negative correlation must also result in the case of precession,
but can deviate and even become positive when changes in $\dot{m}$ are 
invoked. In the absence of precession, the emission in both bands is a
sensitive function of inclination (see \citealt{2007_Poutanen_MNRAS.377.1187P,
2015_Middleton_MNRAS.447.3243M}).

NASA's \textit{Swift} satellite offers an unrivalled opportunity to explore the
long timescale changes in both the X-ray and UV bands through simultaneous
observing by the XRT and UVOT instruments. For a sample of $\sim 40$ ULXs,
we have extracted the UV and X-ray light curves and searched for correlations,
placing constraints on the Pearson coefficient via Monte Carlo simulations.

Our sample contains three Galactic sources, V404 Cygni, Swift J0243.6+6124 and
SS433. Strictly speaking, Swift J0243.6+6124, is a ULX under the
classical empirical definition, however the other two sources are indirectly identified as
supercritical accretors and share many of the same properties of extra-galactic ULXs.
V404 Cygni and Swift J0243.6+6124 display among the strongest positive correlations
in the sample; the strength of these correlations can be ascribed to large outbursts
which subsequently decay over time in both the X-ray and UV/optical bands.
The outbursts from both of these sources have been previously  studied
(see \citealt{2019_Oates_MNRAS.488.4843O} and \citealt{2022_Liu_A&A...666A.110L}),
however, to our knowledge this is the first time UVOT data has been used in a study
for Swift J0243.6+6124. SMC X-3 is another source that displayed an X-ray outburst in 2016
(\citealt{2017_Weng_ApJ...843...69W, 2017_Tsygankov_A&A...605A..39T, 2017_Townsend_MNRAS.471.3878T}); before
dimming below the \textit{Swift} detection threshold, many of the observations
were taken in windowed timing mode without simultaneous observations in the UV. The combination
of the large dynamic range in X-ray count rate and the low number of datapoints may well
bias our inferred correlations for this source.

NGC 300 ULX-1 has a bright counterpart and the system has been identified as a ULX pulsar with a supergiant companion 
\citep{2018_Carpano_MNRAS.476L..45C, 2019_Heida_ApJ...883L..34H}; it is also the closest of the ULXs in our sample to  consistently
display correlated variability. Initially thought to be supernova due to its transient nature in the optical 
\citep{2010_Monard_CBET.2289....1M}, the source has been studied using the
UVOT by \citep{2016_Villar_ApJ...830...11V}. We observe that the B band
appears to display a negative correlation. From the lightcurves (available in the
supplementary material), we observe that the source was bright in the B band around 2010 with
count rates peaking at $\sim 10-15 \mathrm{ct \ s^{-1}}$ and X-ray count rates of 
$\sim 0.05 \mathrm{ct \ s^{-1}}$ (in the  full X-ray band). In 2017 the opposite behaviour occurred
with X-ray count rates peaking at $\sim 0.12 \mathrm{ct \ s^{-1}}$ and B band count rates
of $\sim 4 \mathrm{ct \ s^{-1}}$.

There are clearly several objects where the correlation between bands is
negative. This can result from precession at fixed accretion rate, a changing
accretion rate without precession but at viewer inclinations {\bf not} into the
wind-cone, or from irradiation of the outer disc for viewer inclinations into
the wind-cone. This appears to be the case for
NGC 300 ULX-1 and Holmberg IX X-1 (see bottom left figure \ref{fig:mean_all_corr}), both of which are systems
where precession has been suggested by previous authors
\citep{2018_Weng_ApJ...853..115W, 2019_Vasilopoulos_MNRAS.488.5225V}. It is
also apparent that in some cases, the nature of the correlation changes for a
single source between bands (see Holmberg IX X-1 in figure \ref{fig:correlations_2}. This likely indicates that
each band is affected to a differing degree by one of the different processes mentioned above. 

\begin{figure}
    \centering
    \includegraphics[width=\columnwidth]{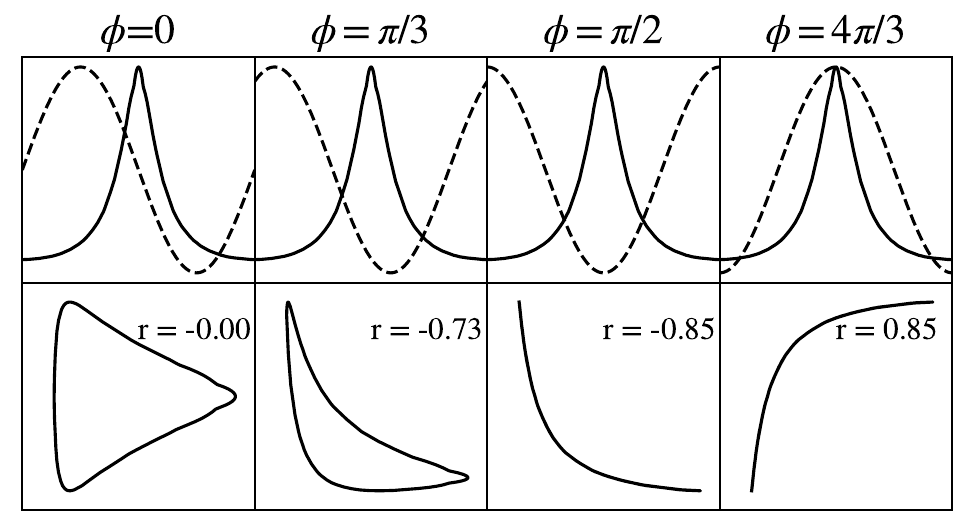}
    \caption[Simulated correlation profiles]{Superimposed light curves with different phases (top row)
    obtained from the the combination of a sigmoid-like light curve (black) from \textsc{ULXLC},
    and a sinusoidal profile (dotted) predicted for the UV emission.
    The bottom row shows the resulting correlation shapes and the associated $r$ value.
    The parameters for the ULXLC used were $\theta = 5^{\circ}$, $\Delta i = 9^{\circ}$ and $i=15^{\circ}$
    (see \citealt{2017_Dauser_MNRAS.466.2236D}).
    }
    \label{fig:ulxlc_profiles}
\end{figure}

\begin{figure*}
    \centering
    \includegraphics[width=0.8\textwidth]{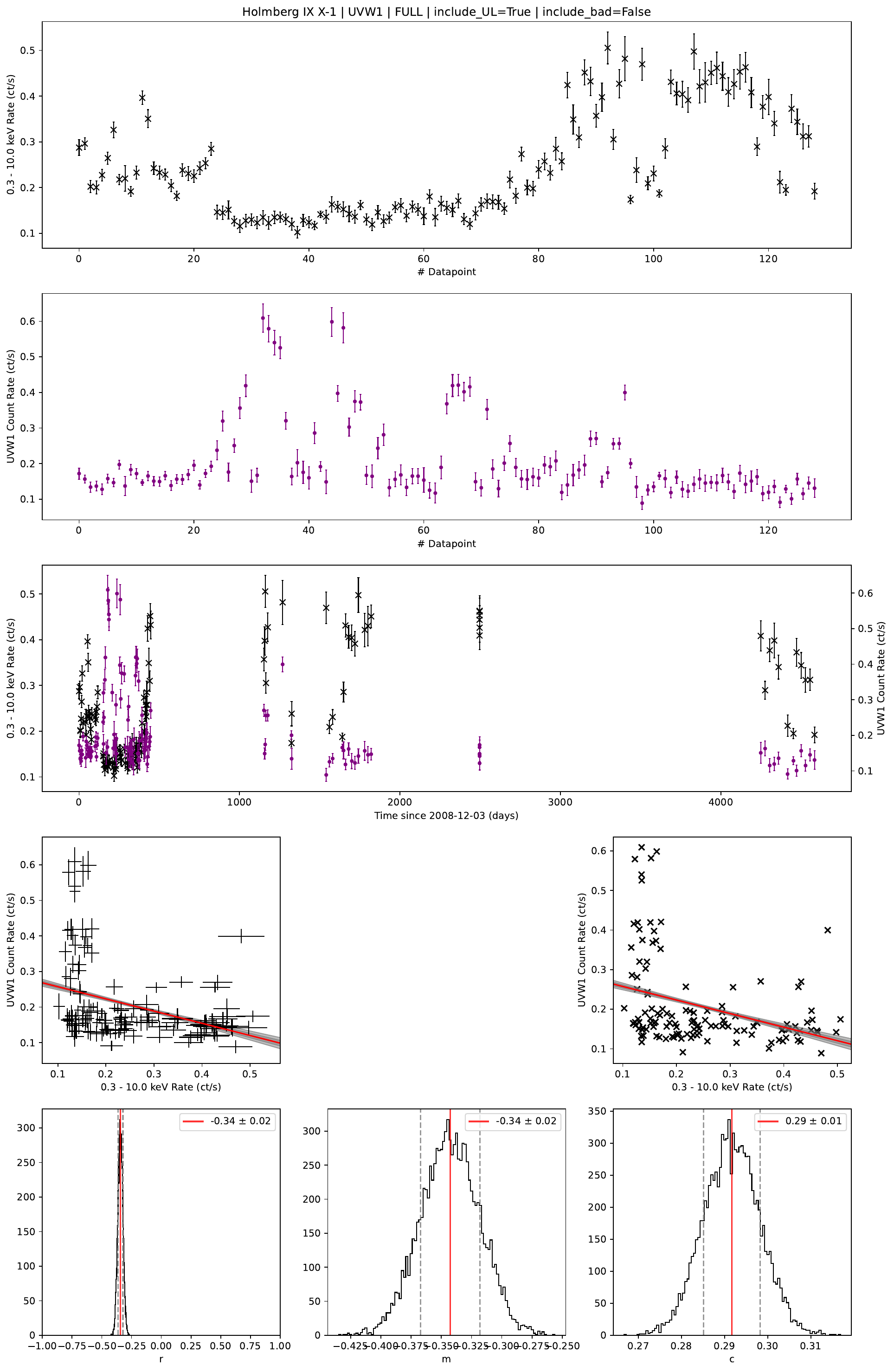}
    \caption{Row 1 \& 2: The \texttt{full} band XRT and UVW1 light
    curves for Holmberg IX X-1 (plotted sequentially) with simultaneous observations.
    Row 3: The same light curves overlaid.
    Row 4: The correlated data points, with and without errors, plotted with the best fitting linear model in red
    (1 and 2$\sigma$ contours are shaded in grey).
    Row 5: The distribution of the fit parameters $r$, $m$ and $c$ for these correlations.}
    \label{fig:Holmberg_IX_lc_fits}
\end{figure*}

In our analysis, we have assumed the simplest case of a linear correlation.
However, numerous sources show clear patterns of behaviour where an `L' shape
is mapped out (see figure \ref{fig:Holmberg_IX_lc_fits}), while others show non-linear
shapes; a simple linear correlation test (Pearson) is naturally less sensitive to
detecting such behaviours. Intriguingly, an `L' shape naturally results from
precession of the wind-cone; using \texttt{ULXLC}
\citep{2017_Dauser_MNRAS.466.2236D} we create an example X-ray lightcurve
using the parameters $\theta = 5^{\circ}$, $\Delta i = 9^{\circ}$ and
$i=15^{\circ}$, which provides a quasi-sigmoid profile (see the first row of figure
\ref{fig:ulxlc_profiles}). We assume that the UV/optical  profile is
sinusoidal (assuming a quasi-spherical geometry of the outer photosphere of the
wind), and plot the subsequent profile of the correlation that would result for
different phases of precession. In the case of $\phi=0$, the two curves produce a triangular shape which results
in a Pearson correlation coefficient of $r=0$, which demonstrates that
even though the two curves are correlated, the value of $r$ does not reveal this. As the phase is changed to $\pi/2$ the curves produce
a characteristic `L' shape which is similar to some of those seen in our light-curves (such
as Holmberg IX X-1).  This preliminary result may
suggest that more complicated (physical) models may be required to understand and search for correlated
variability between bands, and that systems which show such `L' shaped correlations,
may be our clearest indication of precession.

\section*{Acknowledgements}
NK acknowledges support via STFC studentship project reference: 2115300. 

MM is supported by STFC through grant ST/V001000/1.

This work made use of data supplied by the UK Swift Science Data Centre at the University of Leicester.

The authors thank the anonymous referee for useful suggestions.

\section*{Data Availability}
The data underlying this article are available in the supplementary material (online).

The data includes the XRT and UVOT lightcurves, Source tables with extraction regions,
mean count rates and number of observations, the supplementary material also includes the full
set of simulation results in this paper.

The source code for this paper may be found at:
\href{https://github.com/nx1/anticorr_data/}{https://github.com/nx1/anticorr\_data/}\\


\bibliographystyle{mnras}
\bibliography{mnras_template}

%%%%%%%%%%%%%%%%%%%%%%%%%%%%%%%%%%%%%%%%%%%%%%%%%%

%%%%%%%%%%%%%%%%% APPENDICES %%%%%%%%%%%%%%%%%%%%%

\appendix

\section{Irradiated Companion Model} \label{sec:irradiated_companion}

We model a cone of emission centred at the origin $(0,0,0)$ with opening parameter
$c = \mathrm{sin}(\theta / 2)$ where $\theta$ is the full opening angle of the cone.
We then place a companion star, modelled as a sphere placed at $(x_0, y_0, z_0)$ with
radius $r$.

The equations and the cone and sphere in Cartesian coordinates are given by equations \ref{eq:cone} and
\ref{eq:sphere} respectively, this geometry allows us to define the $z$ axis as being in-line with the 
radiation of the beamed cone of emission and adjust the position of the star relative to this axis.

\begin{equation}
    \frac{x^2+y^2}{c^2} = z^2
    \label{eq:cone}
\end{equation}

\begin{equation}
    (x-x_0)^2 + (y-y_0)^2 + (z-z_0)^2 = r^2
    \label{eq:sphere}
\end{equation}

We then solve \ref{eq:cone} and \ref{eq:sphere} simultaneously for $x$, $y$ using the \texttt{SymPy}
Python package \citep{2017_Meurer_10.7717/peerj-cs.103}, this provides four solutions, positive and negative for both $x(z)$ and $y(z)$.

\begin{figure}
    \centering
    \includegraphics{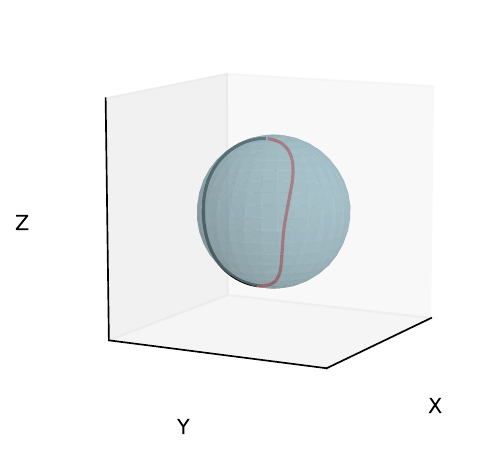}
    \caption{The intersection of a cone with opening angle $\theta = 15^{\circ}$ and a star at position $x_0=2$,
    $y_0=0$, $z_0=10$ and a radius of $r=2.2$. These set of parameters make it such that the cone is grazing the star.
    This figure is an intermediate step in our calculations where we have yet to cull the path due
    to obscuration of the star itself.} 
    \label{fig:3d_intersection_grazing}
\end{figure}

These equations allow us to calculate the $x$ and $y$ coordinates for a given $z$
that corresponds intersection between the cone of emission and the sphere, 
by ignoring the $z$ coordinate and plotting only the $x$ and $y$ values, we may obtain the
projection that would be observed by an observer situated down the
line-of-sight of the $z$ axis. Since the cone is centred at the origin,
by applying a rotation $R_{x}(i)$ or $R_{y}(i)$ and re-projecting down the z axis,
we can simulate the effect of an observation at a specified inclination (with
respect to the jet axis).

We then cut any values of the intersection that are above the hemisphere of the star ($z>z_{0}$)
as these parts would not be illuminated by the beam.
If the configuration is such that the star passes through one wall of the cone as seen
in figure \ref{fig:3d_intersection_grazing}, then we consider this as an edge case and
explicitly calculate the new path by joining the culled path with the arc created from
the circle bounded by the hemisphere of the star.

The result is a closed path, defined by x, y coordinates of the area that would be seen
by an observer, the area is calculated via two methods, one using the shoelace formula
and the other via triangulation both of consistently produce the same result.

%%%%%%%%%%%%%%%%%%%%%%%%%%%%%%%%%%%%%%%%%%%%%%%%%%

% Don't change these lines
\bsp	% typesetting comment
\label{lastpage}
\end{document}